\documentclass[aps,prl,reprint,floatfix]{revtex4-1}

\usepackage{bm}
\usepackage{graphicx}
\usepackage{xcolor}
\usepackage{amsmath,amssymb,physics}
\usepackage{siunitx}
\DeclareSIUnit\Molar{\textsc{m}} 

\renewcommand{\vec}[1]{\bm{#1}}
\newcommand{\un}{\hat{\vec{\mathrm{e}}}}
\newcommand{\unz}{\un_z}

\newcommand{\unr}{\un_r}
\newcommand{\unth}{\un_\theta}

\newcommand{\vdr}{\vec{v}}

\newcommand{\ii}{\mathrm{i}}

\newcommand{\fieldpar}{\xi}
\newcommand{\curvepar}{\delta}

\newcommand{\dip}{^\text{dip}}
\newcommand{\rep}{^\text{rep}}

\renewcommand{\phi}{\varphi}

\newcommand{\ec}{\mathrm{e}}
\newcommand{\kb}{k_\text{B}}
\newcommand{\eps}{\epsilon}

\newcommand{\Kd}{K_\text{d}}

\newcommand{\Du}{\mathrm{Du}}

\newcommand{\jdiff}{\vec{j}_\text{diff}}
\newcommand{\johm}{\vec{j}_E}

\newcommand{\ionpar}{\mu}
\newcommand{\thermalpot}{\phi_\text{th}}
\newcommand{\R}{a}

\newcommand{\tzeta}{\widetilde{\zeta}}
\newcommand{\tsigma}{\widetilde{\sigma}}

\newcommand{\patt}{p}

\newcommand{\dpf}{\vec{h}}
\newcommand{\rdf}{\vec{g}}
\newcommand{\rpf}{\vec{k}}
\newcommand{\pattspeed}{\tau_\patt^{-1}}

\begin{document}

\title{Emergence of colloidal patterns in AC electrical fields}
\date{\today}
\author{Florian Katzmeier}
\author{Bernhard Altaner}
\author{Jonathan List}
\author{Ulrich Gerland}
\author{Friedrich Simmel}

\affiliation{%
Physics Department E14 \& T37, TU Munich, D-85748 Garching, Germany
}%

\begin{abstract}
Suspended microparticles subjected to AC electrical fields collectively organize into band patterns perpendicular to the field direction.
The bands further develop into zigzag shaped patterns, in which the particles are observed to circulate.
We demonstrate that this phenomenon can be observed quite generically by generating such patterns with a wide range of particles: silica spheres, fatty acid, oil, and coacervate droplets, bacteria, and ground coffee.
We show that the phenomenon can be well understood in terms of second order electrokinetic flow, which correctly predicts the hydrodynamic interactions required for the pattern formation process.
Brownian particle simulations based on these interactions accurately recapitulate all of the observed pattern formation and symmetry-breaking events, starting from a homogeneous particle suspension.
The emergence of the formed patterns can be predicted quantitatively within a parameter-free theory.

\begin{description}
\item[PACS numbers]
\end{description}

\end{abstract}

\pacs{Valid PACS appear here}
\maketitle

Systems driven far from equilibrium can self-organize into spatiotemporal dissipative structures and thereby undergo spontaneous symmetry breaking~\cite{Prigogine.1967,Prigogine.1968}.
Such dynamic behavior has been observed in electrokinetic experiments with clay particles~\cite{B.R.JenningsM.Stankiewicz.1990}, polystyrene micro-spheres~\cite{Hu.1994}, and also with $\lambda$-DNA~\cite{Isambert.1997}.
When an alternating electric field is applied, particles form chains along the field direction.
These initially formed particle chains then move towards each other, align in parallel and develop extended band patterns roughly perpendicular to the field direction.
The particle chains within the bands undergo dynamic break-up, 
resulting in the formation of triangular band structures wherein particles are observed to circulate.
Due to the generic occurrence of the band patterns it has been previously asserted that the observed phenomena might be quite general in nature~\cite{Isambert.1997}. 

Originally, Jennings attributed the chain break-up to dipole like repulsion forces arising from electrophoretic particle oscillations~\cite{B.R.JenningsM.Stankiewicz.1990}.
Hu~et~al. explained the particle circulation with  electrorotation caused by mutual polarization of the particles~\cite{Hu.1994}.
Further experimental studies following this interpretation were conducted by  Lele et al.~\cite{Lele.2008} and  Mittal et al. \cite{Mittal.2008}.
For observations with $\lambda$-DNA, Isambert and coworkers assumed that hydrodynamic interactions were generated by local conductivity gradients caused by electrophoretic salt depletion~\cite{Isambert.1997}, resulting in liquid shearing under the influence of an external electric field.
All of the above models explained the dynamics within the band structures, but did not address their formation in the first place.  

In the present work we verify the generic emergence of the same characteristic patterns for a wide range of particles, including bacteria, fatty acid droplets, oil droplets, silica micro-spheres, ground coffee and coacervate droplets. We demonstrate that the observed phenomena can be naturally explained with an electrokinetic fluid flow~\cite{schnitzer2013weaklynonlinear,SQUIRES.2004,GeoffreyIngramTaylor} around colloidal particles, which is also explicitly observed around larger particles. A Brownian particle simulation that specifically accounts for the hydrodynamic and dipole-dipole pair interactions
reproduces the key aspects of the band pattern formation  such as the spontaneous breaking of lateral symmetry, the inclination of the bands with respect to the field direction and the particle circulation within these bands. For the case of the silica spheres, we experimentally investigate the emergence of patterns as a function of 
salt concentration and electric field frequency. The observed dependence can be predicted without any free parameters from the weakly nonlinear multi-scale theory of Schnitzer, Yariv, and coworkers~\cite{schnitzer2012modelbasics,schnitzer2013weaklynonlinear}.


\begin{figure*}[htb!]
\centering
    \includegraphics[scale=0.9]{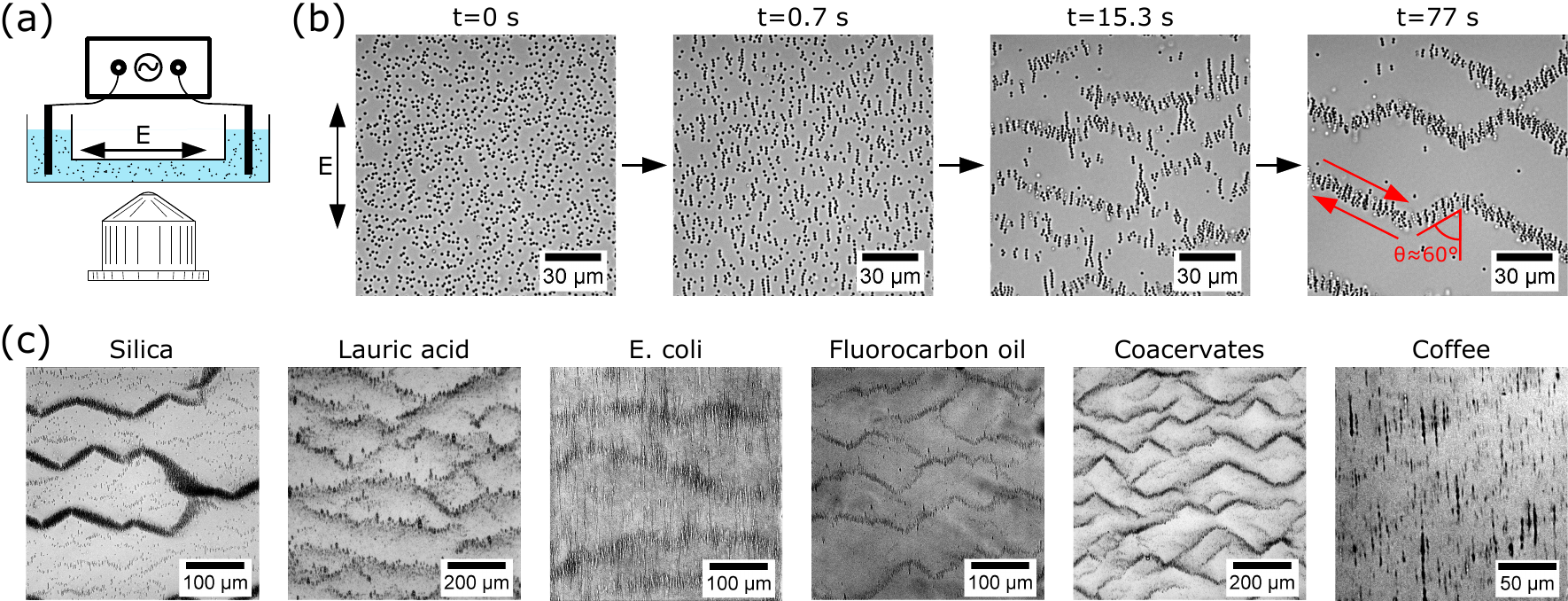}
    \caption{(a) Schematic representation of the experimental setup. Aqueous particle suspensions are subjected to AC electrical fields inside of a microscopic observation chamber.
      (b) Snapshots of a suspension of \SI{1.3}{\micro\meter} diameter silica particles in an AC electric field 
      ($E_0    = \SI{17}{\milli\volt/\micro\meter}$, $f=\SI{500}{\hertz}$) at different timepoints.
      The formation of zigzag-shaped band patterns is clearly visible.
  In the fully formed bands ($t=\SI{77}{\second}$) the particles circulate as indicated.
(c) Band formation and zigzag patterns in suspensions of silica particles, lauric acid droplets, E.coli bacteria, polyallylamine/ATP coacervates, and ground coffee.
The images are taken $\approx\SI{100}{\second}$ after the electric field was turned on }
    \label{fig:experiments}
\end{figure*}

\emph{Experiments.} We conducted our experiments with aqueous suspensions of various micrometer-scale particles, including fluorocarbon (FC) oil and lauric acid droplets, coacervates made from poly(allylamine) and adenosine triphosphate (ATP), monodisperse silica particles (radius~$\R \approx \SI{650}{\nano\meter}$), {\it E.~coli} bacteria, and ground coffee
(for their preparation see chapter 1 of the SI~\cite{SI_text}).
The suspensions were loaded into microscope observation chambers with platinum electrodes placed at opposite inlets (see Fig.~\ref{fig:experiments}(a)).
Fluorocarbon oil droplets, bacteria and coffee particles were imaged in a custom-made glass chamber, while silica particles, lauric acid droplets and coacervates were 
observed in a commercial plastic chamber. After letting the colloids sediment for \SI{10}{\min}, we applied an in-plane AC electric field and recorded the resulting dynamics on the bottom of the chamber with an inverted microscope. 
We applied electric fields between \SI{17}{\milli\volt/\micro\meter} and \SI{56}{\milli\volt/\micro\meter}, which is on the order of the thermal voltage ($\thermalpot := \kb T /\ec \approx \SI{25.69}{\milli\volt}$) for \si{\micro\meter} sized particles. The applied frequency was set to \SI{500}{\hertz} for all samples except for bacteria, where it was \SI{250}{\hertz}.

We found that similar band patterns formed in all of our samples (Fig.~\ref{fig:experiments}).
For all particle types, chain formation occurred within the first second after the electric field was switched on, while horizontal band structures emerged within the first minute.
The band structures continued to grow and merge until the electric field was switched off.
The time course of the pattern formation process is exemplarily shown for silica particles in Fig.~1(b), and can be clearly observed in the supplementary videos~\cite{Fig_1}. 
Snapshots of the patterns taken \SI{100}{\second} after the electric field was turned on are shown in Fig.~\ref{fig:experiments}(c). 
We observed distinct zigzag patterns for coacervates, FC oil and silica particles, and less dominant patterns for the polydisperse lauric acid droplets.
For the bacteria we found only chain formation at $f=\SI{500}{\hertz}$, while at $f=\SI{250}{\hertz}$ the typical band patterns emerged, but no pronounced zigzag structures.
The sample containing the polydisperse ground coffee particles showed more irregular behavior, but
chain formation and also the onset of band formation could be clearly observed. 
The angle between the zigzag bands and the electric field axis was roughly \SI{60}{\degree} (highlighted in Fig.~\ref{fig:experiments}(b)). \\


\emph{Physical mechanisms and theory.}
The initially observed formation of particle chains is well-known~\cite{B.R.JenningsM.Stankiewicz.1990,Hu.1994,Lele.2008,Mittal.2008} and is simply caused by induced dipole-dipole interactions which are usually derived by employing an external AC electric field given as the real part of the complex phasor $\vec{E}(t)= E_0 e^{\ii \omega t} \unz$ with angular frequency $\omega = 2\pi f$. 
The time-averaged dipole-dipole force  on a particle at position $\vec{r}$ exerted by another residing at the coordinate origin is then given as
\begin{align}
  \vec{F}\dip(\vec{r})   & = 6 \pi \eps  \abs{\Kd}^2 E_0^2 \R^2\dpf\left(\vec{r}\right),
  \label{eq:force-on-dipole}
\end{align}
 with the complex dipole coefficient $\Kd$, the permittivity of water $\eps$ and the substitution  $\dpf(\vec{r}) := \frac{1-3\cos^2{\theta}}{r^4}\unr - \frac{2 \cos{\theta} \sin{\theta}} {r^4}\unth$,
where $\theta$ is the zenith angle in spherical coordinates, and $r$ is given in units of the particle radius $\R$.
 
 \begin{figure}[htb]
    \centering
    \includegraphics[width=0.45\textwidth]{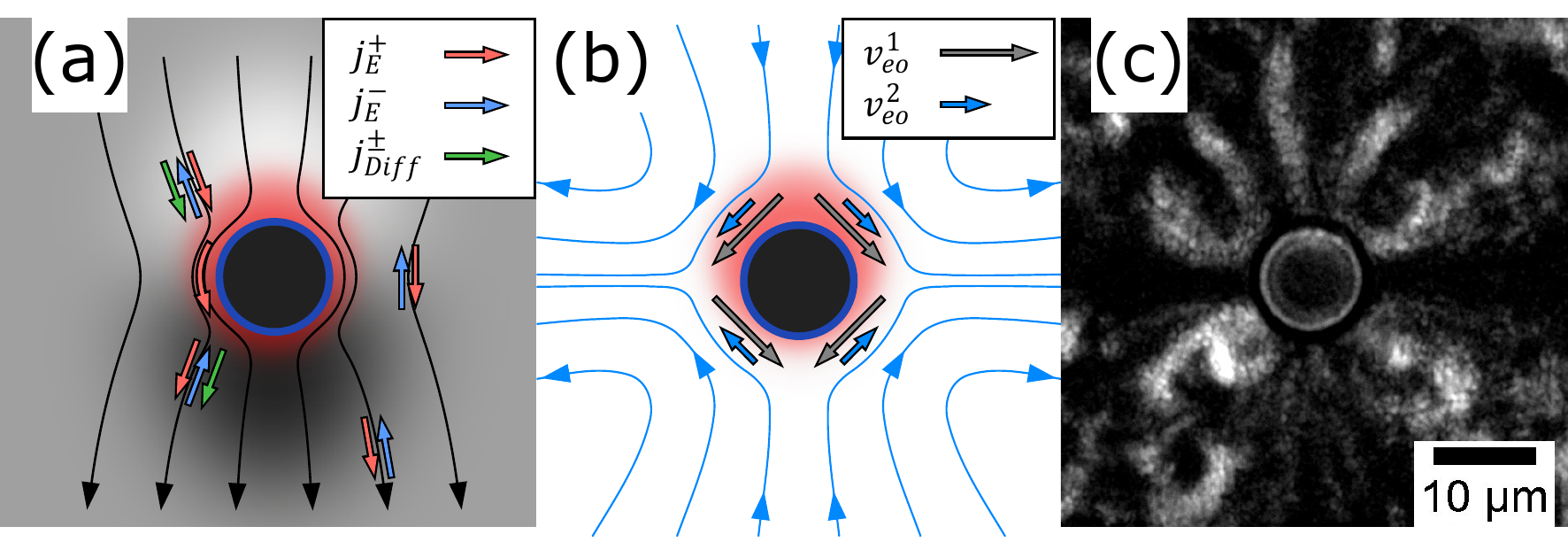}
    \caption{
      (a) A charged particle in an electric field (black field lines) surrounded by counterions (red cloud) and ion fluxes (outlined arrows). The concentration of the neutral salt background is 
      drawn as a grey cloud with darker regions corresponding to higher concentrations. 
(b) Time averaged electrokinetic flow arising from the Coulomb forces acting on the diffuse charge cloud on the particle surface. 
The first and second order contribution to the electrokinetic slip condition are drawn with outlined arrows. 
(c) Experimentally observed fluid motion around a large coacervate (superposition of a \SI{3.8}{\second} long video with enhanced contrast for moving particles).
Bright areas correspond to trajectories of smaller coacervates.
    }
    \label{fig:interaction}
\end{figure}

The formation of the large scale patterns is driven by hydrodynamic interactions, which are caused by electrokinetic flow around the particles. The corresponding
Poisson--Nernst--Planck--Navier--Stokes system of nonlinear partial differential equations can only be solved approximatively \cite{Bikerman.1940,OBrien.1978,OBrien.1981,OBrien.1983,Masliyah.2006,S.S.Dukhin.1965,Wiersema.1966,Ohshima.1983,Ohshima.1995,Ohshima.2006,grosse2012full,grosse2009DC,grosse2009AC}.
For the DC case, Schnitzer, Yariv and coworkers~\cite{schnitzer2012modelbasics,schnitzer2013weaklynonlinear} recently developed a weakly non-linear electrokinetic theory, in which the dimensionless electrokinetic flow $\vec{\tilde{u}}:= \vec{u}/u^*$, with $u^* = \frac{(\kb T)^2 \eps}{\ec^2 \R \eta}$ ,
is expanded in powers of the dimensionless electric field~$\fieldpar =\frac{\ec\R }{\kb T}E_0$:
\begin{align}
  \vec{\tilde{u}} = \fieldpar \vec{\tilde{u}}_1 + \fieldpar^2 \vec{\tilde{u}}_2 + \fieldpar^3 \vec{\tilde{u}}_3 + \dots.
  \label{eq:streaming-function-expansion}
\end{align}
From the DC solution one can extrapolate to the time-averaged AC solution. Here one can use the fact that odd powers of $\fieldpar \propto E_0 e^{\ii \omega t} $ have a zero time average, which leaves $\vec{\tilde{u}}_2$ as the leading order electrokinetic flow. An explicit expression for $\vec{\tilde{u}}_2$ can be deduced from the stream function given in~\cite{schnitzer2013weaklynonlinear}, which results in
\begin{align}
  \vec{u} = \frac{1}{2} u^* \fieldpar^2 \vec{\tilde{u}}_2 = u^* \fieldpar^2 \frac{\gamma}{2}\left(  \rdf -  \dpf \right) 
  \label{eq:average-velocity-leading-order}
  \end{align}
 where  $\rdf(\vec{r}) := \frac{1-3 \cos^2{\theta}}{r^2} \unr$ is a radial field
and $\gamma$ is a dimensionless microscopic parameter (see below and Ref.~\cite{schnitzer2013weaklynonlinear}).
Notably, this well-known Taylor flow pattern~\cite{GeoffreyIngramTaylor,SQUIRES.2004}
can be explicitly observed around larger coacervate droplets through the trajectories of the smaller droplets
(Fig.~\ref{fig:interaction}(c), see also Supporting Video~\cite{Fig_2}), which nicely follow the streamlines of the electrokinetic flow shown in Fig.~\ref{fig:interaction}(b).\\
The derivation of the individual terms in equation~(\ref{eq:streaming-function-expansion}) is quite involved~\cite{schnitzer2012modelbasics,schnitzer2013weaklynonlinear}, but the mechanism
can be understood qualitatively from Fig.~\ref{fig:interaction}: a negatively charged particle immersed in an electrolyte is surrounded by a diffuse charge layer, in which positive counter-ions are accumulated (red cloud), while co-ions are almost completely depleted. Outside the diffuse layer, whose expansion is on the order of the Debye screening length $\kappa^{-1}$ 
the salt solution is electrically neutral.
 The electrokinetic properties of the diffuse layer are determined by the zeta potential $\zeta$ which depends on the surface charge and $\kappa^{-1}$ which in turn depends via $\kappa^{-1} = \sqrt{\frac{2 \ec^2 n}{\eps \kb T}}$ on the number density $n$ of (monovalent) ions.
 The asymmetry in ionic concentrations results in an ion-selective surface conductivity, which is characterized by the ion-selective Dukhin number~$\Du$~\cite{schnitzer2012modelbasics}.

The electric field drives Ohmic counter- and co-ionic currents $\johm^+$ and $\johm^-$, which are aligned with the field lines. Along field lines entering the diffuse layer, 
flux balance requires an additional diffusive current $\jdiff$ which is given by a net salt concentration gradient around the particle that counterbalances the co-ion current $\johm^-$.
Consequently, the neutral salt concentration $n$ varies between the top and the bottom of the colloid
which causes a locally varying perturbation $\zeta_1$ of the equilibrium zeta potential $\zeta_0$ i.e., $\zeta=\zeta_0+\zeta_1$.
This perturbation is visible in Fig.~\ref{fig:interaction}(a) as an expansion of the diffuse layer on one side and a compression on the other side of the particle.

The free charges in the diffuse layer are subject to a Coulomb force due to the tangential component of the electric field $E_\theta$, which
gives rise to fluid motion in the diffuse layer according to the electrokinetic slip condition $v_{eo}= \zeta E_\theta$.
The first and second order velocity components $v_{eo}^1$ and $v_{eo}^2$ connected to $\zeta_0$ and $\zeta_1$ are indicated in Fig.~\ref{fig:interaction}(b).
For alternating electric fields $v_{eo}^1$ has a zero time average, while $v_{eo}^2$ has a non-vanishing time average, resulting in a fluid 
flow around the particle, which is also depicted in the figure.

\begin{figure*}[ht]
  \centering
  \includegraphics[width=1\textwidth]{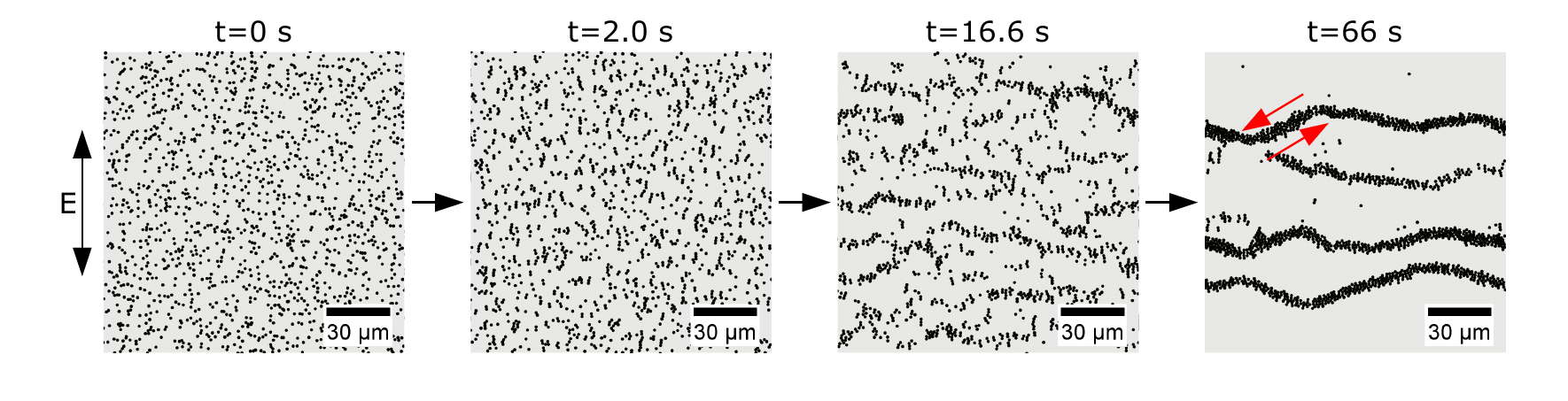}
  \caption{
    Simulated dynamics with $\gamma = 0.088$ and $\abs{\Kd}^2 = 0.23$. Snapshots of a large scale simulation of the patterning process. The different stages of the experimentally observed patterning process from figure~\ref{fig:experiments}(b) are nicely reproduced. The particles circulate again as indicated.
    }
  \label{fig:simulation}
\end{figure*}


\emph{Brownian dynamics simulation.} In the overdamped limit, a direct force $\vec{F}_i^\text{tot}$ exerted on particle $i$ results in particle drift with velocity $\frac{D_p}{\kb T} \vec{F}^\text{tot}_i$ where $D_p$ is the particles diffusion constant.
To include hydrodynamic interactions with the flow field $\vec {u}_i^\text{tot}(\vec{r})$, which is caused by other particles ($j \neq i$) in the fluid, we use
Faxen's correction for the drift velocity: 
\begin{align}
\vdr_i= \frac{D_p}{\kb T}\vec{F}^\text{tot}_i + \left(1+ \frac{1}{6}\Delta \right)\vec{u}_i^\text{tot}. 
  \label{eq:drift-with-hydrodynamics}
\end{align}

The direct force on particle $i$ is obtained as the sum of dipolar and a repulsive interactions $\vec{F}\rep(\vec{r}_{ij})$  i.e. $
  \vec{F}_i^\text{tot}=\sum_{j \neq i} \left( \vec{F}\dip(\vec{r}_{ij}) + \vec{F}\rep(\vec{r}_{ij})\right),
$
where $\vec{r}_{ij} = \vec{r}_i -\vec{r}_j$ denotes the difference vector between particles $i$ and $j$. 
Ignoring geometric interactions, the velocity field caused by the particles $j \neq i$ is to zeroth order given as the sum
$
  \vec{u}^\text{tot}_i = \sum_{j \neq i} \vec{u}(\vec{r}_{ij}).
$
Together with \eqref{eq:force-on-dipole} and \eqref{eq:average-velocity-leading-order}  and by recognizing that $\Delta \rdf = -6 \dpf$ and $\Delta \dpf =0$, the drift velocity (Eq.\eqref{eq:drift-with-hydrodynamics}) becomes
\begin{align}
  \vdr_i(\vec{r}_i) = \nonumber
  u^* \sum_{j\neq i} 
  &\bigg[ \fieldpar^2\left( \frac{\gamma}{2} \rdf(\vec{r}_{ij}) + (\abs{\Kd}^2 - \gamma) \dpf(\vec{r}_{ij})\right) \nonumber \\
  &+ \nu \rpf(\vec{r}_{ij}) \bigg] 
  \label{eq:final-form}
\end{align}
where the repulsion $\nu \rpf(\vec{r}_{ij})$ is discussed in chapter 3 of the SI~\cite{SI_text}.
The movement of the particles can then be described by 
the  $N$-particle Langevin equation, 
\begin{align}
  \R \dd {\vec{r}}_i  = \vdr_i \dd{t}  + \sqrt{2D_p} \dd{W}_i,
  \label{eq:smoluchowski-equation}
\end{align}
where  $\dd{W}$ is the stochastic increment of a Wiener process.

We solved this stochastic differential equation SDE using the Euler-Maruyama algorithm with periodic boundary conditions and random initial particle configurations. (cf. chapter~3 of the SI~\cite{SI_text})
As we observed the emergence of stripe patterns exclusively at the channel bottom, we restricted our simulation to two dimensions 
by constraining the 3D expressions of the dipole-dipole force and fluid flow to the plane $y=0$.
Despite this simplification we capture both geometry and scaling of the physical interactions at least qualitatively correctly. 
As shown in Fig.~\ref{fig:simulation}, a simulation based on Eq.~\eqref{eq:smoluchowski-equation} with 1521 particles, $\gamma = 0.088$, $\abs{\Kd}^2 = 0.23$, and a particle density matched to our silica particle experiments correctly recapitulates all stages of the observed pattern formation process (cf. Fig.~\ref{fig:experiments} and the SI Video~\cite{Fig_3}). 

\emph{Parameter dependence of the stripe patterns.} To gain further insight into the physical mechanisms underlying the pattern formation process, we explored its dependence on AC field frequency and salt concentration. To this end, we prepared aqueous suspensions of monodisperse silica particles at 0.0375~\%~(w/v) with NaCl concentrations ranging from \SI{5}{\micro\Molar} to \SI{2500}{\micro\Molar}, 
and recorded microscopy videos with a relatively weak electric field amplitude of \SI{10.6}{\milli\volt\per\micro\meter} at frequencies ranging from \SI{250}{\hertz} to \SI{25}{\kilo\hertz}.

To analyze our data, we defined the `pattern visibility' $\patt$ in an image as
the discretized version of
$ \patt=  \int_A  |\frac{\partial}{\partial z} \left( G \ast I \right) (x,z) | dx dz$
where $A$ is the area of the image, $G$ is a Gaussian function with a standard deviation of 15 pixels (corresponding to \SI{6}{\micro\meter}), $I(x,z)$ is the intensity of the image, and $\ast$ denotes convolution. The defined order parameter $\patt(t)$ is time-dependent and measures density fluctuations along the $z$-direction at a scale defined by $G$.

\begin{figure}[h]
  \centering
  \includegraphics[width=.4\textwidth]{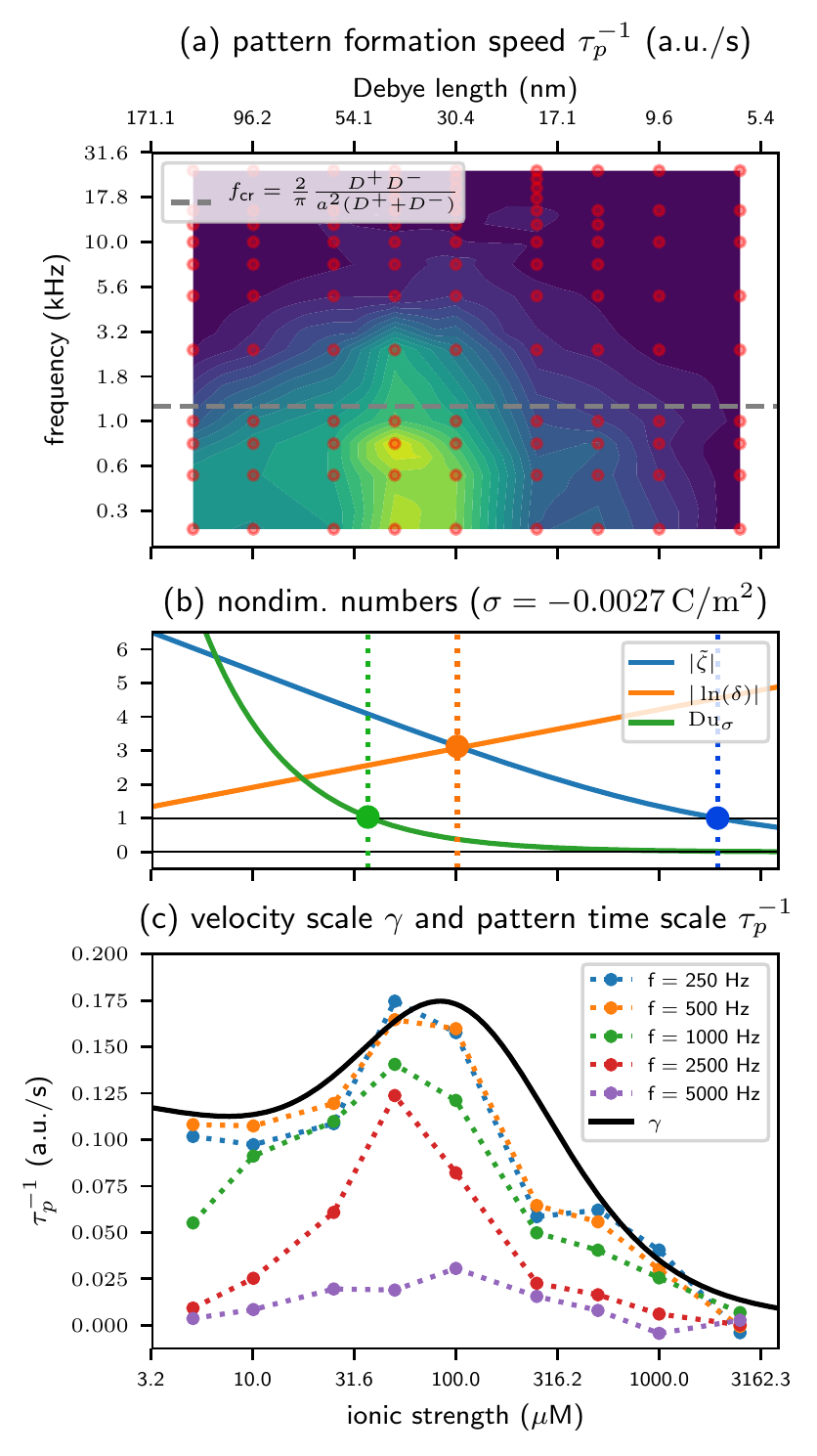}
  \caption{
    (a) Observed pattern visibility $\patt$ in an experiment.
    Red circles indicate the experimental data points with a heat map generated by linear interpolation on the logarithmic grid.
    (b) Dimensionless characteristic numbers for electrokinetics lead to characteristic scales for the ionic strength (or, equivalently, Debye length), see main text for details.
    (c) The low-frequency pattern visibility $\patt$ can be quantitatively predicted without fit parameters using the second-order velocity scale $\gamma$ derived by Schnitzer and Yariv in Ref.~\cite{schnitzer2013weaklynonlinear}.
  }
  \label{fig:quantitative}
\end{figure}

We computed $\patt(t)$ for every frame of our microscopy videos and used it to determine a
typical timescale $\tau_p$ for the emergence of the stripe patterns (example curves of $\patt(t)$ are shown in SI Chapter~2~\cite{SI_text})
In Fig.~\ref{fig:quantitative}(a),  $\pattspeed$ is plotted as a measure for the speed of the pattern formation process for various values of the field frequency and salt concentration.


\emph{Microscopic theory and dimensionless parameters.} As we identified electrokinetic fluid flow as the driving mechanism behind the pattern formation process, we expect $\pattspeed$ to scale with the magnitude of the fluid flow, which is set by $\gamma$.
The microscopic parameter $\gamma$ is related to the physics of the Debye layer, whose details are specified by the curvature parameter $\curvepar := (\kappa a)^{-1}$ and the dimensionless equilibrium zeta potential $\tzeta_0=\zeta_0/\thermalpot$~\cite{schnitzer2012modelbasics,schnitzer2013weaklynonlinear}.
The latter is linked to the dimensionless surface charge density $\tsigma :=  \frac{\sigma}{ \eps \kappa \thermalpot} $ 
by the Grahame equation $ \tsigma = 2 \sinh{\tzeta_0/2}.$\\
The ionic transport around colloidal particles is characterized by the Dukhin number $\Du$, which measures the relative strength of surface to bulk conductivity~\cite{Dukhin.1991,delgado2005report,dukhin1993dukhinnumber}. By considering the surface conductivity of counter-ions only, an ion-selective Dukhin number~\cite{schnitzer2012modelbasics,schnitzer2013weaklynonlinear} given by
 $\Du_\sigma := \curvepar\tsigma(1+2\ionpar^+)$ can be defined with the ionic drag coefficient $\ionpar^+ := \frac{\eps \thermalpot^2}{\eta D^+}$ and the counter-ion diffusion constant $D^+$\cite{schnitzer2012modelbasics,schnitzer2013weaklynonlinear}.
In Figure \ref{fig:quantitative}(b), we show the variation of the relevant dimensionless numbers for the ionic conditions of our experiment, where we set the surface charge
of the silica particles 
to the known value $\sigma = \SI{-0.0027}{\coulomb\per\meter^2}$~\cite{shi2018surfacecharge}.
Comparison with Fig.~$\ref{fig:quantitative}(a)$ indicates that patterns can be observed only up to a characteristic ionic strength where $\abs{\tzeta_0}$ (blue dot) 
is $O(1)$. For higher ionic strengths, $\abs{\zeta_0} < \thermalpot$, the physics of the Debye layer can be neglected altogether. Further, pattern formation is fastest in the parameter range where the zeta potential is `logarithmically large' compared to the curvature parameter i.e. $\abs{\tzeta} =  O(\abs{\ln(\curvepar)})$ (orange dot) and where surface conduction becomes dominant over bulk conduction i.e. $\Du_\sigma = O(1)$ (green dot).

Finally, in Fig.~\ref{fig:quantitative}(c) we compare  $\gamma$ (Ref.~\cite{schnitzer2013weaklynonlinear}) with the observed pattern visibility $\patt$,
for which we scaled $p$ such that its maximum at $f= \SI{250}{\hertz}$
corresponds to the maximum value of $\gamma$. 
For the lowest experimental frequencies, we find excellent agreement between pattern visibility $\patt$ and $\gamma$, and even 
for higher frequencies $p$ qualitatively shows the same behavior, albeit with a reduced amplitude. Notably, the start of the pattern formation process for salt concentrations below \SI{1}{\milli \Molar} as well as the maximum pattern formation speed at around \SI{50}{ \micro \Molar} are nicely predicted by $\gamma$, when the known value of $\sigma = \SI{-0.0027}{\coulomb\per\meter^2}$~\cite{shi2018surfacecharge} is used. Further, we find that value of $\gamma$ chosen for our simulation (Fig.~\ref{fig:simulation}) has a physically reasonable magnitude.
While no nonlinear AC theory is available to date, we find that the decrease of the amplitude falls in the range of the characteristic frequency $f_\text{cr}=\frac{2}{\pi}\frac{D^+D^-}{a^2(D^++D^-)}=\SI{1.2}{\kilo \hertz}$ of the diffusive cloud, which is known from other linear AC theories~\cite{shilov2001review,Lyklema.1983,Lyklema.1986,Grosse.1996,grosse2009AC,grosse2012full}. 

In conclusion, we experimentally verified the generic occurrence of a pattern formation process that had been previously observed when different types of colloids in aqueous suspension were subjected 
to AC electrical fields. 
We identified the physical mechanisms underlying the pattern forming process as dipole-dipole interactions and second order electrokinetic fluid flow, and confirmed the 
emergence of collective behavior in a many particle simulation. We found that Schnitzer-Yariv's weakly non-linear electrokinetic theory gives a parameter-free quantitative explanation of the pattern formation process, which only requires a surface charge on the colloidal particles, providing a satisfactory unifying explanation for the observed macroscopic patterns and their underlying physical mechanism. 

Apart from its fundamental scientific interest, the described effect could be utilized for various applications in microfluidics and microrobotics, in which AC fields generally are 
advantageous as they cause fewer Faradaic processes.
For instance, it should be possible to use the described electrokinetic flow for the implementation of microfluidic pumps and mixers. Similar devices based on induced charge electroosmosis have been the focus of intense research efforts in the past~(e.g.,~\cite{Bazant.2006,Hong.2011,Ng.2009,Huang.2010,Urbanski.2006,Wu.2008}). Further, we believe that our insights will be helpful for the development of electrically manipulated microswimmers. While such microswimmer systems were previously envisioned~\cite{SQUIRES.2006} and implemented~\cite{Gangwal.2008,Peng.2014,Mano.2017,Nishiguchi.2015,Nishiguchi.2018,Boymelgreen.2015} based on inorganic (metallo-dielectric) Janus particles, our experimental results with oil droplets, coacervates, lauric acid and even bacteria demonstrate that microswimmers consisting solely of soft and biological material are feasible. 

\paragraph{Acknowledgments}
This work was funded by the Deutsche Forschungsgemeinschaft (DFG, German Research Foundation) – Project-ID 364653263 – TRR 235, and the European Research Council 
(grant agreement no. 694410 - project AEDNA).
Jonathan List gratefully acknowledges support through a stipend by the Peter und Traudl Engelhorn Stiftung. 

\bibliography{references}

\cleardoublepage
\newpage

\end{document}